\documentclass[a4paper,11pt]{article}

\usepackage{jinstpub} 
\usepackage{tabulary}
\usepackage{placeins} 
\usepackage{lineno}
\usepackage{caption}
\usepackage{subcaption}
\usepackage{amsmath}
\usepackage{amsthm} 
\usepackage{bm} 
\usepackage{xspace}
\usepackage{upgreek}
\usepackage{float}
\usepackage{booktabs}  
\usepackage[svgnames]{xcolor}

\newcommand{\minerva}{MINERvA\xspace}

\title{Vertex finding in neutrino-nucleus interaction : A Model Architecture Comparison}

\newcommand{\Florida}{University of Florida, Department of Physics, Gainesville, FL 32611}

\newcommand{\CBPF}{Centro Brasileiro de Pesquisas F\'{i}sicas, Rua Dr. Xavier Sigaud 150, Urca, Rio de Janeiro, Rio de Janeiro, 22290-180, Brazil}
\newcommand{\PUCP}{Secci\'{o}n F\'{i}sica, Departamento de Ciencias, Pontificia Universidad Cat\'{o}lica del Per\'{u}, Apartado 1761, Lima, Per\'{u}}

\newcommand{\Pittsburgh}{Department of Physics and Astronomy, University of Pittsburgh, Pittsburgh, Pennsylvania 15260, USA}
\newcommand{\Guanajuato}{Campus Le\'{o}n y Campus Guanajuato, Universidad de Guanajuato, Lascurain de Retana No. 5, Colonia Centro, Guanajuato 36000, Guanajuato M\'{e}xico.}

\newcommand{\Tufts}{Physics Department, Tufts University, Medford, Massachusetts 02155, USA}
\newcommand{\WM}{Department of Physics, William \& Mary, Williamsburg, Virginia 23187, USA}
\newcommand{\FNAL}{Fermi National Accelerator Laboratory, Batavia, Illinois 60510, USA}

\newcommand{\MCLA}{Massachusetts College of Liberal Arts, 375 Church Street, North Adams, MA 01247}
\newcommand{\UMD}{Department of Physics, University of Minnesota -- Duluth, Duluth, Minnesota 55812, USA}

\newcommand{\Rochester}{Department of Physics and Astronomy, University of Rochester, Rochester, New York 14627 USA}

\newcommand{\USM}{Departamento de F\'{i}sica, Universidad T\'{e}cnica Federico Santa Mar\'{i}a, Avenida Espa\~{n}a 1680 Casilla 110-V, Valpara\'{i}so, Chile}
\newcommand{\Geneva}{University of Geneva, 1211 Geneva 4, Switzerland}

\newcommand{\OregonState}{Department of Physics, Oregon State University, Corvallis, Oregon 97331, USA}
\newcommand{\oxford}{Oxford University, Department of Physics, Oxford, OX1 3PJ United Kingdom}

\newcommand{\upenn}{Department of Physics and Astronomy, University of Pennsylvania, Philadelphia, PA 19104}
\newcommand{\AMU}{AMU Campus, Aligarh, Uttar Pradesh 202001, India}

\newcommand{\Mohali}{Department of Physical Sciences, IISER Mohali, Knowledge City, SAS Nagar, Mohali - 140306, Punjab, India}

\newcommand{\york}{York University, Department of Physics and Astronomy, Toronto, Ontario, M3J 1P3 Canada}
\newcommand{\ND}{Department of Physics, University of Notre Dame, Notre Dame, Indiana 46556, USA}
\newcommand{\ICL}{The Blackett Laboratory,  Imperial College London,  London SW7 2BW, United Kingdom}
\newcommand{\warwick}{Coventry CV4 7AL, UK}
\newcommand{\ornl} {Oak Ridge National Laboratory, Oak Ridge, Tennessee 37831, USA}

\newcommand{\nAMU}{a}
\newcommand{\nUSM}{b}
\newcommand{\nCBPF}{c}
\newcommand{\nornl}{d}
\newcommand{\nWM}{e}
\newcommand{\nPUCP}{f}
\newcommand{\nRochester}{g}
\newcommand{\nGuanajuato}{h}
\newcommand{\nGeneva}{i}
\newcommand{\nOregonState}{j}
\newcommand{\nND}{k}
\newcommand{\nUMD}{l}
\newcommand{\nyork}{m}
\newcommand{\nFNAL}{n}
\newcommand{\nMohali}{o}
\newcommand{\nICL}{p}
\newcommand{\nupenn}{q}
\newcommand{\nwarwick}{r}
\newcommand{\noxford}{s}
\newcommand{\nMCLA}{t}
\newcommand{\nTufts}{u}
\newcommand{\nPittsburgh}{v}
\newcommand{\nFlorida}{w}
\author[\nAMU]{F.~Akbar}
\author[\nUSM,\nCBPF]{A.~Ghosh}
\author[\nornl]{S. Young}
\author[\nAMU]{S.~Akhter}
\author[\nWM,\nAMU]{Z.~~Ahmad~Dar}
\author[\nAMU]{V.~Ansari}
\author[\nPUCP]{M.~V.~Ascencio}
\author[\nAMU]{M.~Sajjad~Athar}
\author[\nRochester]{A.~Bodek}
\author[\nGuanajuato]{J.~L.~Bonilla}
\author[\nGeneva]{A.~Bravar}
\author[\nRochester]{H.~Budd}
\author[\nCBPF]{G.~Caceres}
\author[\nRochester]{T.~Cai}
\author[\nOregonState,\nCBPF]{M.F.~Carneiro}
\author[\nRochester]{G.A.~D\'{i}az~}
\author[\nGuanajuato]{J.~Felix}
\author[\nND]{L.~Fields}
\author[\nWM]{A.~Filkins}
\author[\nRochester]{R.~Fine}
\author[\nAMU]{P.K.Gaur}
\author[\nUMD]{R.~Gran}
\author[\nyork,\nFNAL]{D.A.~Harris}
\author[\nFNAL]{D.~Jena}
\author[\nMohali]{S.~Jena}
\author[\nRochester]{J.~Kleykamp}
\author[\nICL]{A.~Klustov\'{a}}
\author[\nupenn]{D.~Last}
\author[\nCBPF]{A.~Lozano}
\author[\nwarwick,\noxford]{X.-G.~Lu}
\author[\nMCLA]{E.~Maher}
\author[\nRochester]{S.~Manly}
\author[\nTufts]{W.A.~Mann}
\author[\nRochester]{K.S.~McFarland}
\author[\nPittsburgh]{B.~Messerly}
\author[\nUSM]{J.~Miller}
\author[\nWM,\nGuanajuato]{O.~Moreno}
\author[\nFNAL]{J.G.~Morf\'{i}n}
\author[\nWM]{J.K.~Nelson}
\author[\nFlorida]{C.~Nguyen}
\author[\nRochester]{A.~Olivier}
\author[\nPittsburgh]{V.~Paolone}
\author[\nFNAL,\nRochester]{G.N.~Perdue}
\author[\noxford]{K.-J.~Plows}
\author[\nupenn,\nGuanajuato]{M.A.~Ram\'{i}rez}
\author[\nRochester]{D.~Ruterbories}
\author[\nPittsburgh]{H.~Su}
\author[\nTufts]{V.S.~Syrotenko}
\author[\nICL]{A.V.~Waldron}
\author[\nUSM]{B.~Yaeggy}
\author[\nWM]{L.~Zazueta}
\affiliation[\nAMU]{\AMU}
\affiliation[\nUSM]{\USM}
\affiliation[\nCBPF]{\CBPF}
\affiliation[\nornl]{\ornl}
\affiliation[\nWM]{\WM}
\affiliation[\nPUCP]{\PUCP}
\affiliation[\nRochester]{\Rochester}
\affiliation[\nGuanajuato]{\Guanajuato}
\affiliation[\nGeneva]{\Geneva}
\affiliation[\nOregonState]{\OregonState}
\affiliation[\nND]{\ND}
\affiliation[\nUMD]{\UMD}
\affiliation[\nyork]{\york}
\affiliation[\nFNAL]{\FNAL}
\affiliation[\nMohali]{\Mohali}
\affiliation[\nICL]{\ICL}
\affiliation[\nupenn]{\upenn}
\affiliation[\nwarwick]{\warwick}
\affiliation[\noxford]{\oxford}
\affiliation[\nMCLA]{\MCLA}
\affiliation[\nTufts]{\Tufts}
\affiliation[\nPittsburgh]{\Pittsburgh}
\affiliation[\nFlorida]{\Florida}

\collaboration{The MINER$\nu$A Collaboration}

\abstract{We compare different neural network architectures for Machine Learning (ML) algorithms
designed to identify the neutrino interaction vertex position in the MINERvA detector. The
architectures developed and optimized by hand are compared with the architectures developed in an automated
way using the package "Multi-node Evolutionary Neural Networks for Deep Learning" (MENNDL),
developed at Oak Ridge National Laboratory (ORNL). The two architectures resulted in a similar performance which suggests that
the systematics associated with the optimized network architecture are small. Furthermore, we find that while the domain
expert hand-tuned network was the best performer, the differences were negligible and the auto-
generated networks performed well. There is always a trade-off between human, and computer resources
for network optimization and this work suggests that automated optimization, assuming resources are
available, provides a compelling way to save significant expert time.
}
 \keywords{
 neutrino, reconstruction, convolutional neural networks, deep learning, 
 }

\begin{document}
\topmargin=0mm 

\maketitle
\section{Introduction}
\label{sec:introduction}
Particle physics experiments are increasing their  
use of Machine Learning (ML) algorithms 
at reconstruction level to maximize their physics output.  These algorithms currently tend to use one specific architecture, which means coming up with values for the number of layers of each type and the number of nodes in each of these layers. The architecture is chosen "by hand", and then the algorithms are trained on both simulated and real particle physics data in the cases of supervised learning and unsupervised learning, respectively. The process of choosing a specific architecture is lengthy and often takes months of human effort.  
Though deep learning has been able to bypass the process of manual feature engineering by learning representations in conjunction with statistical models in an end-to-end fashion, neural network architectures themselves are typically designed by experts in a painstaking, ad hoc manner. Neural architecture search (NAS) has been boosted as the path forward for alleviating this pain by automatically identifying architectures that are superior to hand-designed ones.\\
The idea of automatically learning and evolving network topologies was first explored in~\cite{6790655}. Recently, the pioneering work by~\cite{DBLP:journals/corr/ZophL16} and~\cite{DBLP:journals/corr/BakerGNR16} have led to a number of better, faster and cost-efficient NAS methods, thus attracting a lot of attention to this field.\\
A group from ORNL developed a 
package named MENNDL~\cite{young2015optimizing,young2017evolving} which addresses the model selection problem and eases the demands on data researchers. MENNDL leverages a large number of compute nodes, which  communicate over Message Passing Interface (MPI) to distribute the task of finding the optimal 
architecture
across the nodes of a supercomputer. 
In our previous paper~\cite{Perdue:2018ihs}, we applied ML for improving the vertex reconstruction. We developed the network inside the collaboration for the neutrino events. The successful application of ML brings the following questions; is the developed algorithm  the optimal solution? Can the same architecture be used for similar kinds of datasets but generating from  antineutino events? Do we need a systematic to cover the architecture choice?
In order to investigate these questions, we performed a thorough comparison  between an algorithm that was tuned by hand over several months of researcher's time given an intimate knowledge of a detector's performance and an algorithm that was tuned using MENNDL over a few hours with a supercomputer at Oak Ridge. Exploring many topologies for a specific physics problem may
help us to understand the systematic related questions mentioned above. The performances were compared using simulated antineutrino interactions in the \minerva Detector. 
These interactions provide a useful arena for a model comparison because there is a detailed hit-level simulation of the detector and an important output that is needed for future physics analyses: the spatial location of the neutrino interaction point (or "vertex") in a complex detector geometry.  The comparison involves testing several different models using the MENNDL package for vertex finding, where those models were optimized over different architectures. 
Although automated neural architecture searches are widely used in industry, to the best of our knowledge this is the first time one is being used to analyze neutrino/antineutrino physics data. \\
The paper is organized as follows: In Sec.~\ref{sec:detector}, we describe the \minerva detector. Next in Sec.~\ref{sec:vtx_rec}, we discuss the challenges in reconstructing the vertex of the neutrino interaction. In Sec.~\ref{sec:artisanal_ntwk}, we summarize the network topology of the artisanal model developed by the MINERvA collaboration. In Sec.~\ref{sec:NAS}, we discucss about NAS, followed by Sec.~\ref{sec:MENNDL} where details of MENNDL are described.
%
%
%
In Sec.~\ref{sec:sample_data} we discuss the details of the simulation, and in Sec.~\ref{sec:Result} we discuss the results. Sec.~\ref{sec:conclusion} provides the conclusion.

\section{\minerva dectector}
\label{sec:detector}
\begin{figure}[h!]
\centering
\begin{subfigure}{0.45\textwidth}
  \centering
  \includegraphics[width=1.0\linewidth]{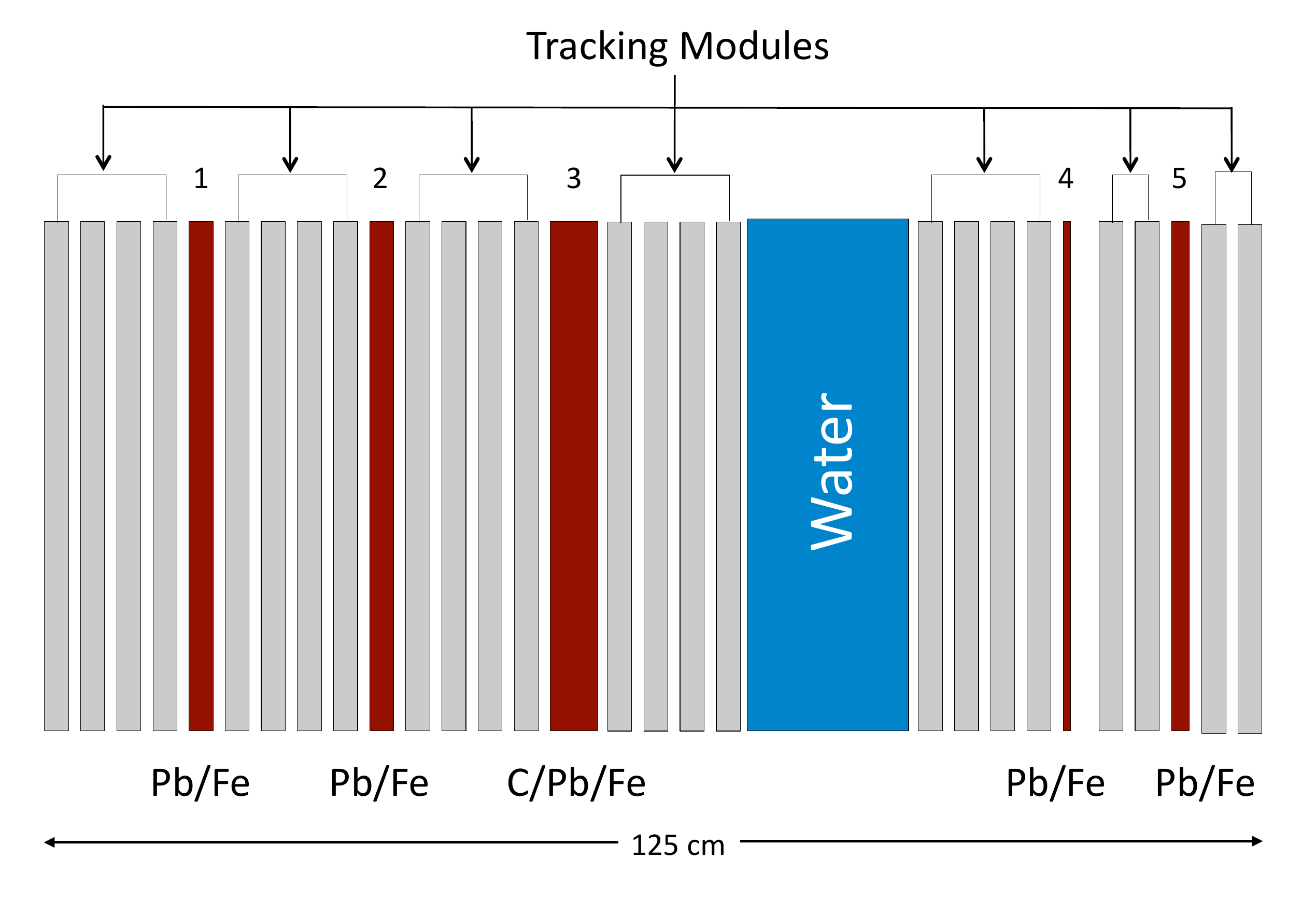}
  \caption{}
  \label{fig:tgt-geo_a}
\end{subfigure}%
\begin{subfigure}{.55\textwidth}
  \centering
  \includegraphics[width=1.0\linewidth]{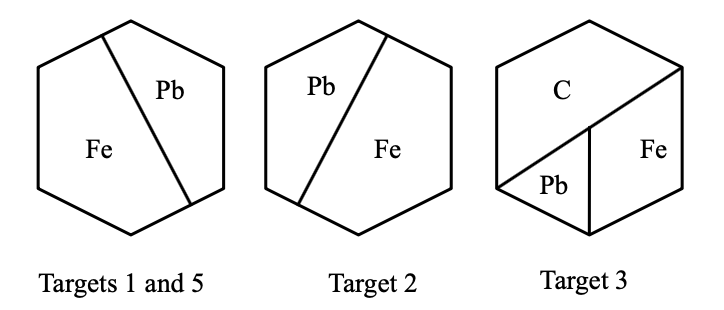}
  \caption{}
  \label{fig:tgt-geo_b}
\end{subfigure}
\caption{ (a) Longitudinal segmentation (b) Transverse segmentation: 
targets 1, 2, 3, and 5 are shown, and target 4 consists entirely of lead.
Target 3 is indicated by the figure at the right, and targets labeled 1, 2, and 5 on the left alternate between the configurations in the left and middle of the diagram at the right (figures taken from \cite{Aliaga:2013uqz}). }
\end{figure}

The \minerva detector~\cite{Aliaga:2013uqz} consists of a nuclear target region where most of the interactions considered in this paper take place, followed by a fine-grained tracker which measures the products from the upstream interactions and also serves as an active target.  There is electromagnetic and hadronic calorimetery surrounding the nuclear targets and the tracking region, and a muon spectrometer (the MINOS near detector~\cite{Michael:2008bc}) is located downstream of the hadronic calorimeter region. 
The nuclear target region of the detector is complex and consists of both active and passive elements.  The region is designed to have the same target material (iron or lead) located in several different regions relative to the center of the neutrino beam and relative to the muon spectrometer which is itself highly non-symmetric with respect to the neutrino beam axis.  The nuclear target design allows cross-checks of the cross section measurements between regions that have different geometric acceptances.  The longitudinal segmentation of the nuclear target region is shown in Fig.~\ref{fig:tgt-geo_a}, where the neutrino beam is incident from the left of the figure.  The transverse segmentation of the different passive nuclear targets is shown in Fig.~\ref{fig:tgt-geo_b}.  

\begin{figure}[h!]
\centering
  \includegraphics[width=1.0\linewidth]{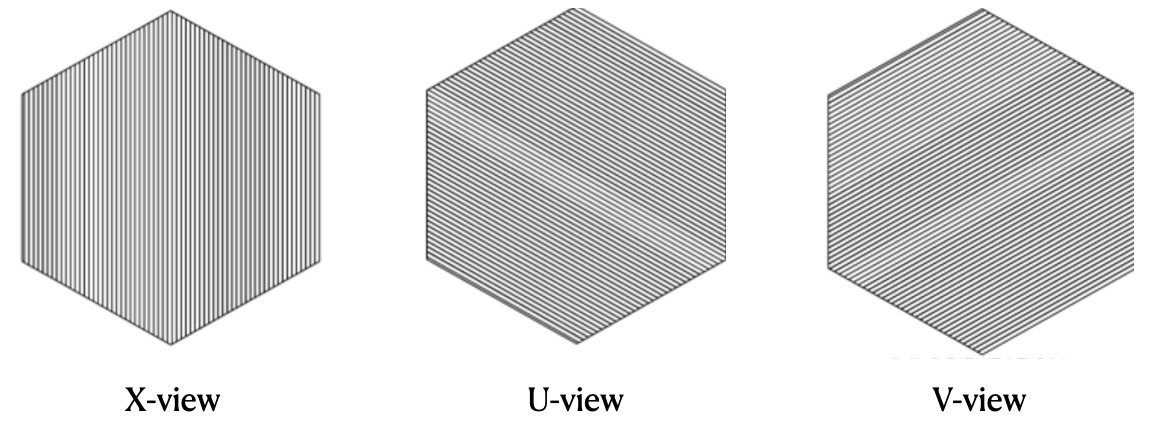}
  \caption{Three three orientation/views of the scintillator planes. }
  \label{fig:views}
\end{figure}

The active scintillator planes that provide the input to the ML algorithm consist of nested bars with a triangular cross section whose base is 3.3 cm and height is 1.7 cm.  The transverse cross section of the detector is hexagonal in shape, and the planes have three different orientations to allow for stereoscopic track reconstruction:  scintillator bars are either vertical or they are oriented $\pm 60^\circ$ with respect to vertical.  Two successive scintillator planes have strips oriented vertically followed by one of either positive or negative $60^\circ$(Fig.~\ref{fig:views}).  Each grey rectangle in Fig.~\ref{fig:tgt-geo_a} corresponds to two scintillator planes where the sequential non-vertical bar directions alternate between $+60^\circ$ and $-60^\circ$.  

\section{Analysis Challenges in the \minerva Experiment}
\label{sec:vtx_rec}
The \minerva experiment has a broad physics program designed to measure a wide range of neutrino interaction channels. One of the primary themes in \minerva's physics program is to measure how different neutrino interaction channels are modified by the nuclear environment where an interaction happens.  In order to do that, we use a detector that has several different thin (passive) targets made of the nuclei we want to study (for example carbon, iron, lead), and then we need to determine the target in which neutrino interaction took place. In order to determine the interaction location, segmented scintillator planes are located both upstream and downstream of the thin targets. The signals from the scintillator planes are then used to predict the start of the interaction.
Neutrinos are neutral particles and leave no trace in the scintillator. Nevertheless, charged particles produced by the neutrino interactions do leave traces, emerging from the interaction vertex in all directions.
In general the higher the neutrino energy, the more energy can be transferred to the nucleus, and the more energy that is transferred, the more final state charged particles can be produced. 
This sounds like a straightforward problem that might not require a ML algorithm, but for high-momentum transfers, there can be several charged particles leaving the interaction point in many directions.  
Vertex finding is made challenging by vertex occlusion in passive material, particle re-interactions, and hadronic shower cascades - these factors combine to significantly degrade vertex finding accuracy, requiring very strict sample selection cuts and background  subtraction estimation procedures to accurately predict and subtract the large background.  We can do better with ML-based reconstruction algorithms.
In a previous publication~\cite{Perdue:2018ihs} we applied Convolutional Neural Network (CNN) based supervised ML model to find the vertex. We considered specific events with high momentum transfers or in other words,  events originating from deep inelastic scattering (DIS) events. Using the ML approach significantly improves the efficiency and accuracy of our vertex-finding algorithms. The CNN-based method was the top performer, including in comparisons with a tracking algorithm based on fitting clusters of hits with a Kalman filter~\cite{10.5555/897831}.
In addition, we presented means to mitigate the possible model biases coming from the large labeled training sample using the Domain Adversarial Neural Network (DANN)~\cite{JMLR:v17:15-239} algorithm.
\section{Network topology of the domain-expert designed ML model }
\label{sec:artisanal_ntwk}

\begin{figure}[ht]
  \centering
  \includegraphics[width=0.7\linewidth]{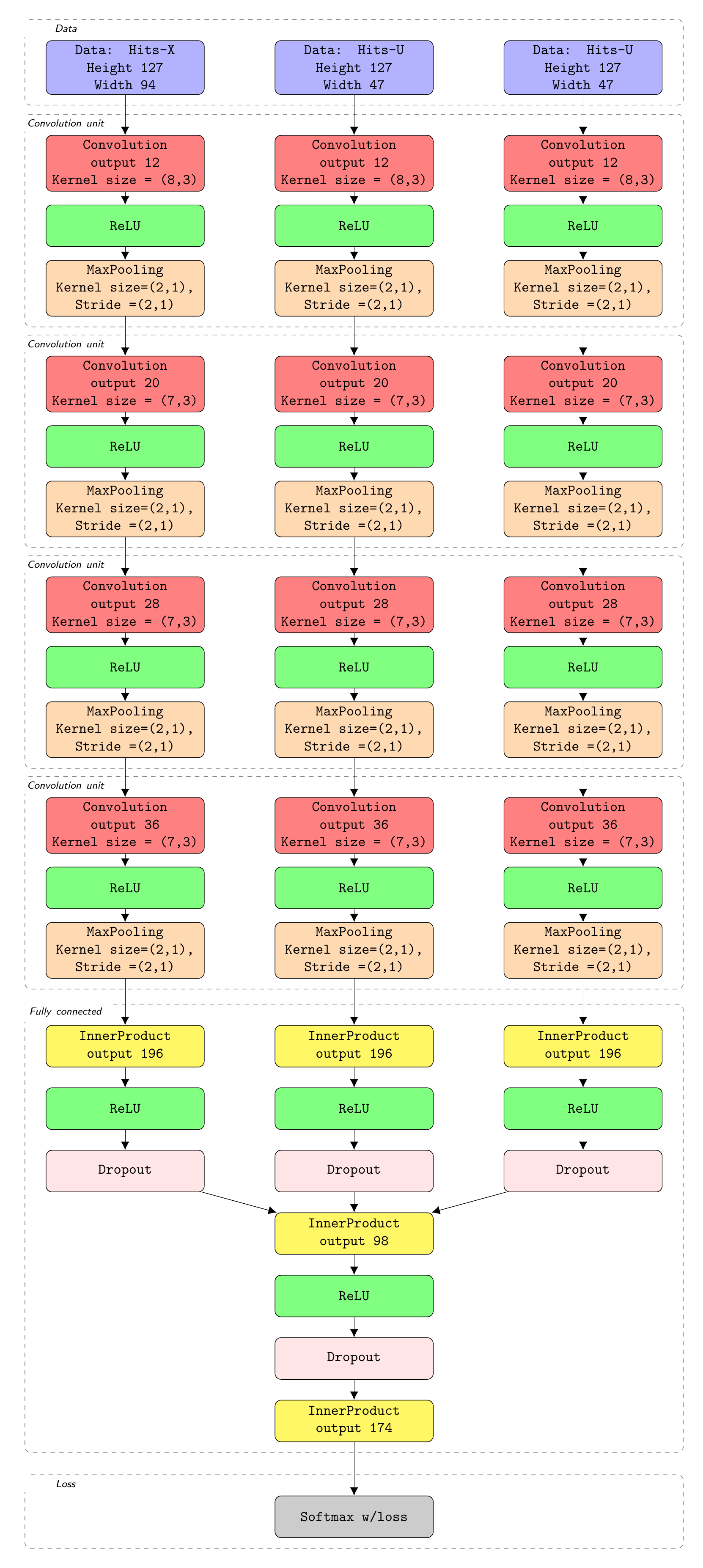}
\caption{The flow chart of the network structure of the artisanal models. The structure of the Artisanal-caffe and Artisanal-Tensorflow are quite similar.}.
  \label{fig:artisanal_network}
\end{figure}

Two CNN based supervised models using Caffe~\cite{jia2014caffe}and Tensorflow~\cite{tensorflow2015-whitepaper} application programming interfaces (APIs) were developed.
The detailed structure of the domain-expert designed, or "artisinal",
Caffe and Tensorflow models are described in Ref.~\cite{Perdue:2018ihs}. 
These two models are indicated as artisanal-Caffe and artisanal-Tensorflow throughout the paper. Both the models are quite similar in structure as for each case, the network is made of three separate towers for X, U and V views where each tower comprises four iterations of convolution and max pooling~\cite{LeCun1989HandwrittenDR}\cite{726791} layers with ReLUs~(Rectified Linear Unit)~\cite{10.5555/3104322.3104425} acting as the non-linear activations. Each pooling layer consists of a kernel which decreases the dimension along the transverse axis by one. After the four iterations of convolution, ReLU and pooling, there is a fully connected layer with 196 semantic output features. The outputs for the three views are concatenated and fed to another fully connected layer with 98 outputs which in turn is input for a final fully connected layer with 67 outputs for the plane classifier, 
which is the input to a Softmax~\cite{Goodfellow-et-al-2016} layer. Fully connected layer allows non-linear combinations of the discovered features and operates at the end of the feature discovery layers to associate the discovered features to desired outputs. 
Finally, we assign the network cost using a cross-entropy function~\cite{pml1Book}, which is subsequently minimized. Fig.~\ref{fig:artisanal_network} shows the structure of the artisnal model diagrammatically. Note that, though the artisanal-Tensorflow model and the aritsanal-Caffe model have the same network structure, there are internal differences, for example in regularization, between two API. Additionally, there are slight differences between the two APIs in the implementation of Stochastic Gradient Descent(SGD)~\cite{Bottou98on-linelearning}. These differences creates slight differences in the model's performance as shown later in the Sec.~\ref{sec:Result} \\
Here, the plane classifier identifies the plane where the true vertex is located. To classify the data with the plane classifier, we use a total of 67 ``planecodes". This includes two ``overflow" classes - code 0 for events upstream of the detector, and code 66 for
everything downstream of the last considered plane. 

\section{Neural architecture searches (NAS)}
\label{sec:NAS}
NAS~\cite{elsken2019neural, DBLP:journals/corr/abs-1902-07638, kasim2020building}
describes a family of automated architecture optimization algorithms
(i.e. number and type of layers,
hyperparameters of those layers, connectivity of layers, etc.). It is distinct from hyperparameter optimization which is generally restricted to optimizer algorithm choice and minor changes to layer hyperparameters of a fixed network architecture. In other words, architecture forms a scaffolding for a neural network and hyperparameters tend to be simple scalar adjustments or algorithm choices given an architecture.
There can be several components of NAS contributing to the whole process. Among them, the most well-defined components are  search space, search strategy, and model evaluation strategy. 
Search space constructs the networks that are further tested to process the final architecture. 
The search strategy determines the method of optimization in order to explore the search space, which significantly affects the efficiency of the search, as well as the effectiveness of the final proposed architecture. 
The choice of the method of optimization ensures the sufficient investigation of the search space and keeps the resulting architecture close to the optimum.
Finally, the evaluation strategy compares the intermediate results, thereby helping the search strategy to choose the best option during the search process.
Several techniques can be employed for the search strategies, e.g., Bayesian optimization~\cite{ru2021interpretable}, evolutionary algorithms~\cite{DBLP:journals/corr/abs-1711-00436}, gradient descent~\cite{9461192}, reinforcement learning~\cite{DBLP:journals/corr/ZophL16}, etc.
\section{MENNDL}
\label{sec:MENNDL}

Multi-node Evolutionary Neural Networks for Deep Learning (MENNDL)~\cite{young2015optimizing,young2017evolving} is a NAS 
algorithm that utilizes
evolutionary optimization for evolving the architecture and hyperparameters of convolutional neural networks on high performance computers.

MENNDL uses a master-worker paradigm such that a master process is used to perform the core evolutionary process and workers are used to evaluate the fitness of the generated networks.
The evolutionary process is asynchronous.
Once a sufficient number of individual networks 
has been evaluated, a new set of individuals are produced through selection, crossover, and mutation and then queued to be evaluated.
This ensures the computational resources used by the workers are not left idle.
MENNDL has previously shown success in automatically designing neural networks for several scientific datasets~\cite{young2015optimizing,young2017evolving}. It is particularly useful for datasets that have very different image characteristics than the photographic imagery typically used in machine learning literature (e.g. ImageNet~\cite{deng2009imagenet}). A key difference between MENNDL and other NAS methods is that it focuses on exploring all possible hyperparameters for the potential layers along with the composition of those layers, as opposed to most other methods which limit the range of hyperparameters to a small set (e.g. DARTS only allows for a handful of potential kernel sizes \cite{liu2018darts}). 

\begin{figure}[ht]
  \centering
  \includegraphics[width=0.8\textwidth]{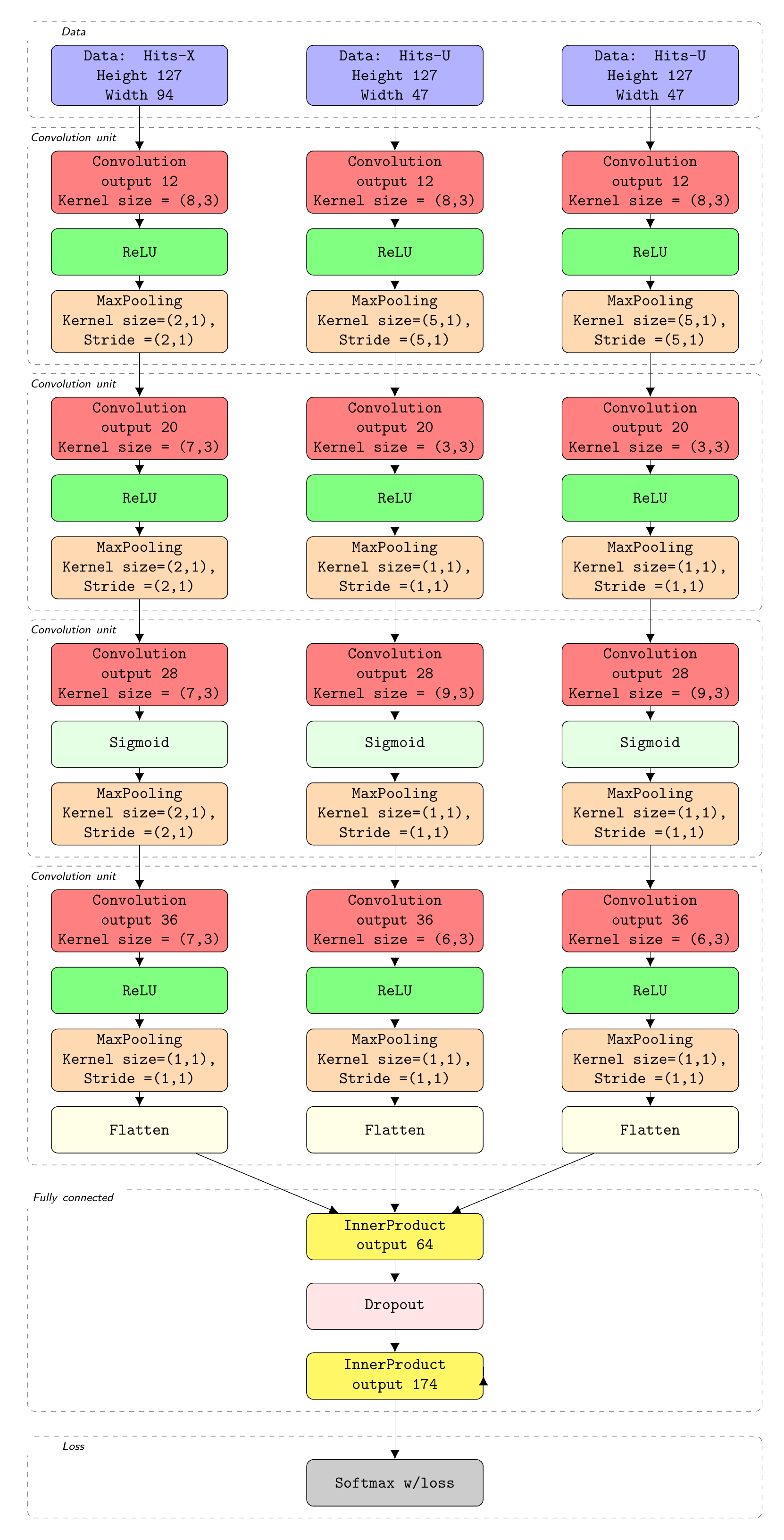}
\caption{The flow chart of the network structure of the MENNDL model (MENNDL 68525 indicated by yellow line in Fig.~\ref{fig:acc_loss_comb_menndl}).  This model performed best among all the MENNDL models. The performance of this model is compared with the artisanal models. }.
  \label{fig:MENNDL_network}
\end{figure}

In this work, some modifications were made in order to handle the unique data used. In order to handle the three views, the same three tower structure was used for the MENNDL networks, with each tower having identical layer types and hyperparameters learned through the evolutionary process. The maximum size of hyperparameters related to the width of the input image was set to the maximum allowed by the two smaller views. MENNDL generated and evaluated approximately $10,000$ networks on the Titan Supercomputer using $500$ nodes for $12$ hours. Example networks are shown in Fig.~\ref{fig:MENNDL_network}. From these networks, a subset was chosen for further training and evaluation.\\
The events used for training and developing the MENNDL networks, are the simulated events generated using neutrino flux peaked around 6 GeV.\\
It is noted that these models are primarily optimized over short run periods (only 5000 iterations) using the neural network software library Caffe.
The version of MENNDL used in this work is defined in the 2017 paper~\cite{young2017evolving}, and thus the hyperparameters optimized by MENNDL in this work are limited to those defining the layers and the architecture of the neural network (e.g. layer types, number of layers, kernel sizes, etc.). Solver hyperparameters such as learning rate are not optimized by MENNDL.

\section{Details of simulation}
\label{sec:sample_data}
Among 10,000 MENNDL models,  we chose nine models by looking 
at their validation accuracy. Validation loss and accuracy are the loss and accuracy on validation set which gives a measure of the quality of the model. 
Next, we considered data sample of size 1.66 million events, and divide that sample in two: $\sim$ 93\% of total sample events are used for training whereas rest ($\sim$ 7\%) are used for validation of the models. These events are the antineutrino events coming from the NuMI beam with a peak energy of around 6 GeV. We trained the MENNDL models and the artisanal models using those simulated antineutrino events up to 20 iterations (epochs). Note that, since we use a CNN-based model, we make images of the input events. A detailed description of image processing is given in Ref.~\cite{Perdue:2018ihs}.  
For studying the impact of the models on a physics analysis, we use another set of simulated data sample of size two million events. 
\section{Results}
\label{sec:Result}
\begin{table}[h]
\begin{center}
\resizebox{\columnwidth}{!}{
\begin{tabular}{|l|l|l|l|l|l|ll}
\cline{1-5}
MENNDL Models            & Total number of parameters & Time of execution  & Validation loss  after 20 epochs & Validation accuracy after 20 epochs (\%)     \\ \cline{1-5}
  MENNDL 57             &  415104             &  15h 36m 37s  & 0.82 & 71.7 \\ \cline{1-5}
  MENNDL 574            &  949248             &  23h 09m 54s  & 0.87 & 70.1 \\ \cline{1-5}
  MENNDL 589833         &  2857568            &  16h 46m 17s  & 1.00 & 67.9 \\ \cline{1-5}  
  MENNDL 633167         &  2857920            &  20h 37m 23s  & 0.85 & 68.5 \\ \cline{1-5}
  MENNDL 68525          &  1637712            &  20h 42m 47s  & 0.73 & 74.4 \\ \cline{1-5}
  MENNDL 68965          &  1637712            &  20h 45m 33s  & 0.74 & 74.2 \\ \cline{1-5}
  MENNDL 982201-1       &  2014224            &  20h 27m 50s  & 0.85 & 70.0 \\ \cline{1-5}  
  MENNDL 982201-2       &  3549240            &  20h 27m 20s  & 0.94 & 66.2 \\ \cline{1-5}
  MENNDL 982201-3       &  2691616            &  20h 21m 35s  & 0.86 & 70.6 \\ \cline{1-5}             
\end{tabular}
}
\end{center}
\caption{ Total number of parameters and time of execution for all the caffe models in Fig.\ref{fig:acc_loss_comb_menndl}. }
\label{param_time}
\end{table}
\begin{table}[h]
\resizebox{\columnwidth}{!}{
\begin{tabular}{|l|l|l|l|ll}
\cline{1-4}
Models               & Validation loss  after 20 epochs & Validation accuracy after 20 epochs (\%) &  Total number of parameters &  \\ \cline{1-4}
Artisanal-Tensorflow & 0.73                               & 76.2                                & 12115716 &  \\ \cline{1-4}
Artisanal-caffe      & 0.78                               & 72.6                                  & 12115716 &  \\ \cline{1-4}
MENNDL (68525)               & 0.73                              & 74.5                                 & 1637712 &  \\ \cline{1-4}
\end{tabular}
}
\caption{ Validation loss and accuracy numbers for each models along with the total number of adjustable parameters in each model.}
\label{acc_num}
\end{table}

\begin{figure}[h!]
\centering
\begin{subfigure}{0.5\textwidth}
  \centering
  \includegraphics[width=1.0\linewidth]{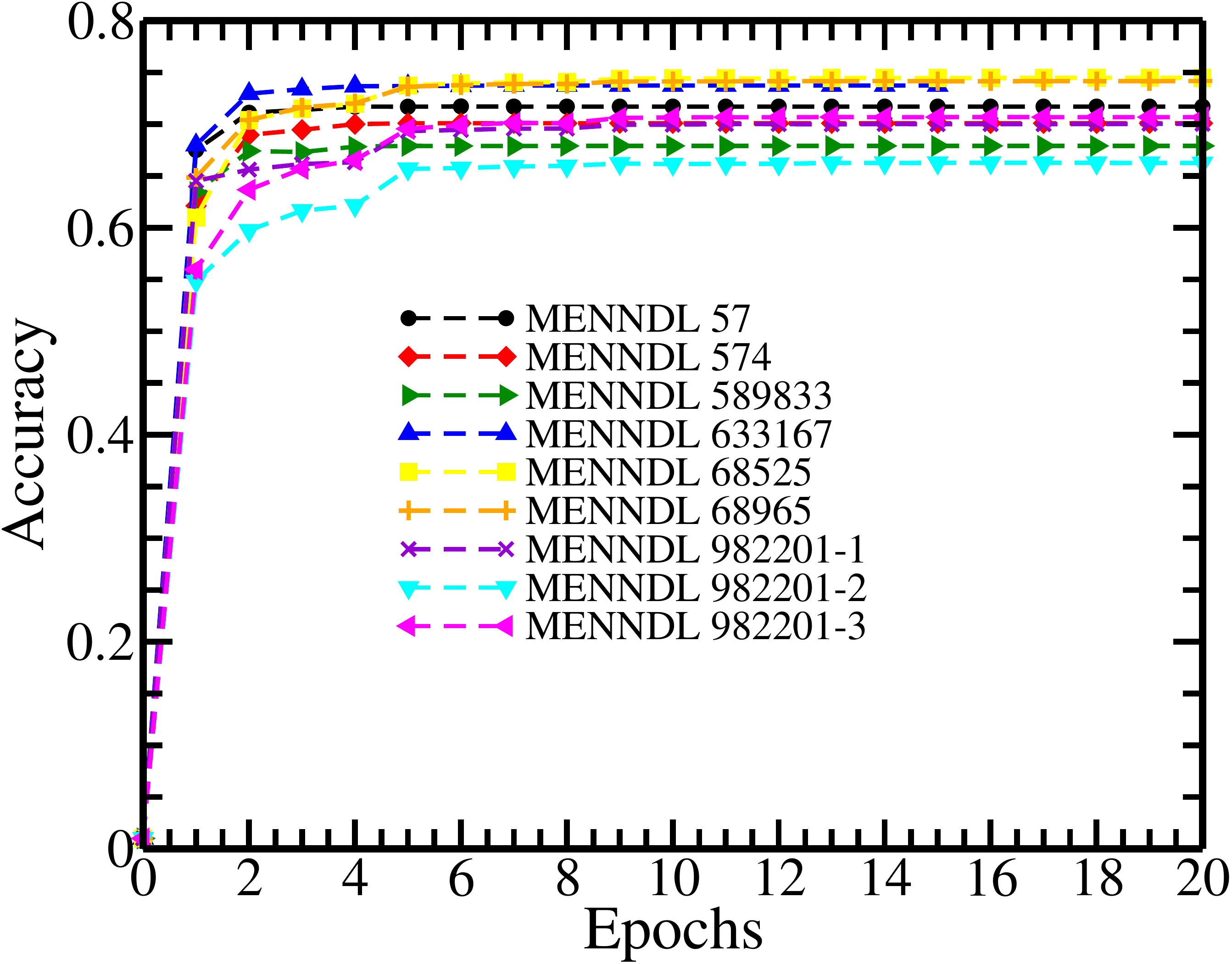}
  \caption{Validation accuracy}
  \label{fig:acc_comb_menndl}
\end{subfigure}%
\begin{subfigure}{.5\textwidth}
  \centering
  \includegraphics[width=1.0\linewidth]{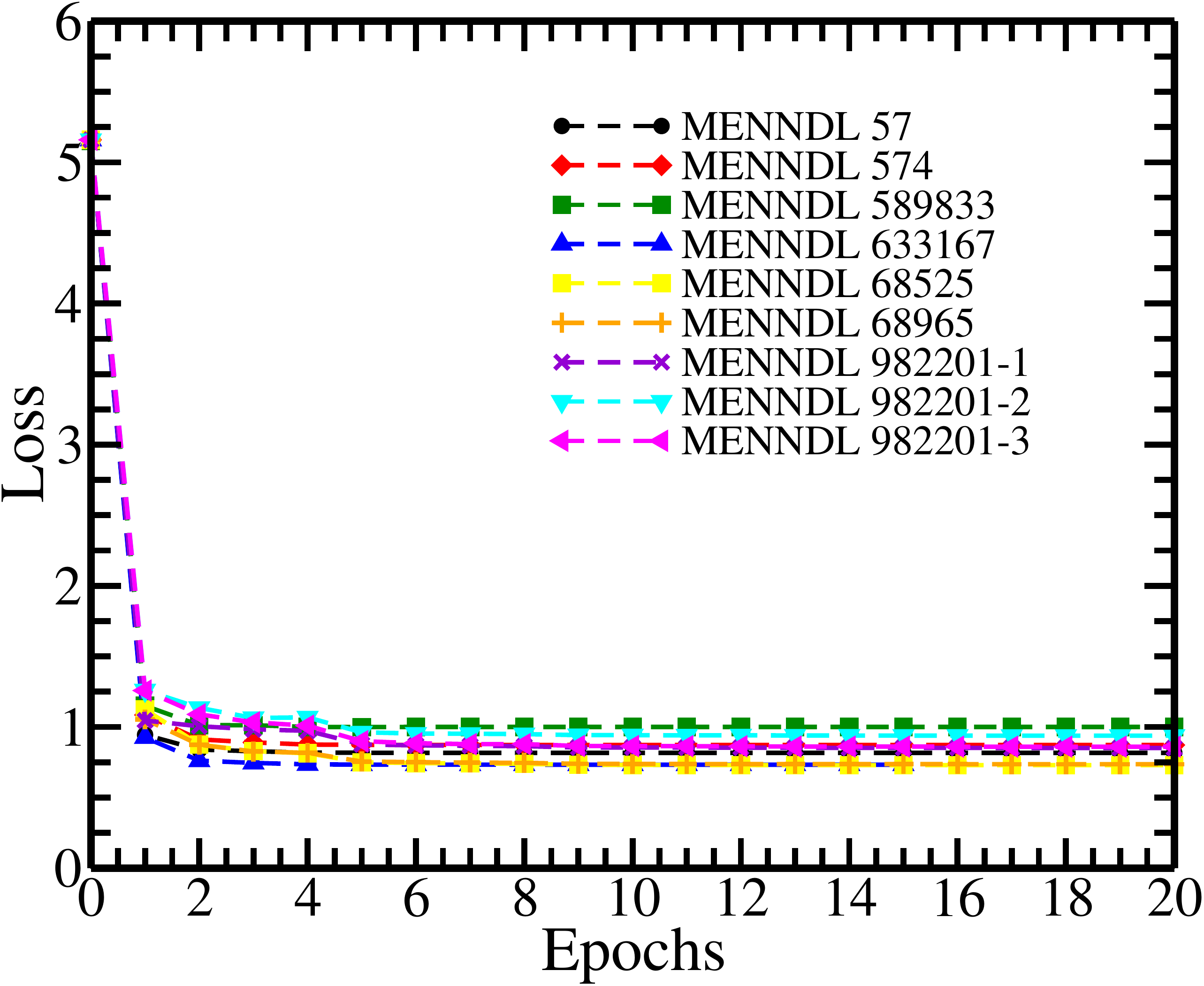}
  \caption{Validation loss}
  \label{fig:loss_comb_menndl}
\end{subfigure}
\caption{ Validation accuracy and validation loss for different MENNDL models. Among these nine models, model {\it MENNDL 68525} (yellow coloured square-dashed line) has the highest accuracy (74.5\%) and lowest 
loss (0.73) values.}.
\label{fig:acc_loss_comb_menndl}
\end{figure}

\begin{figure}[h!]
\centering
\begin{subfigure}{0.51\textwidth}
  \centering
  \includegraphics[width=1.0\linewidth]{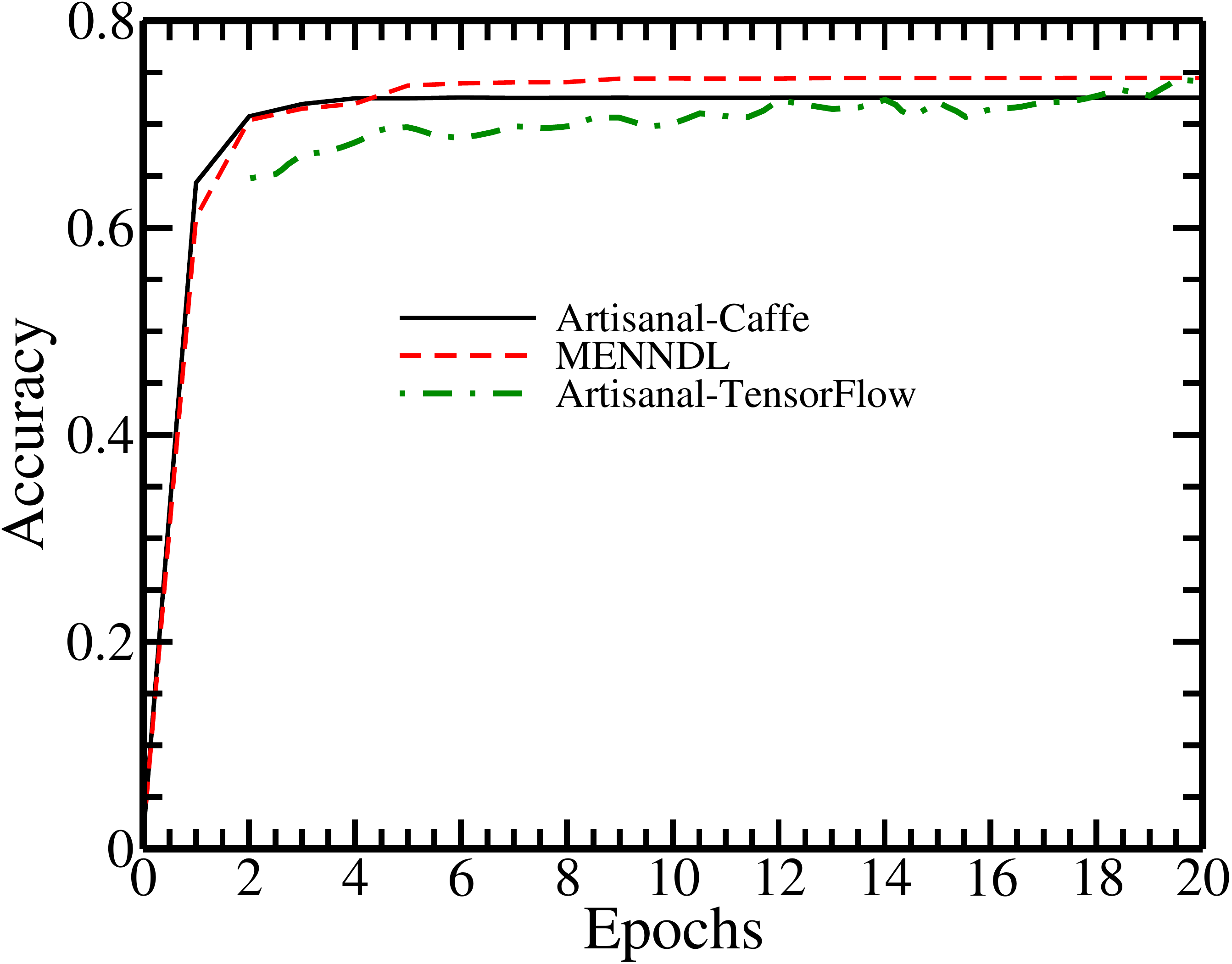}
  \caption{Validation accuracy}
  \label{fig:acc_comp}
\end{subfigure}%
\begin{subfigure}{.49\textwidth}
  \centering
  \includegraphics[width=1.0\linewidth]{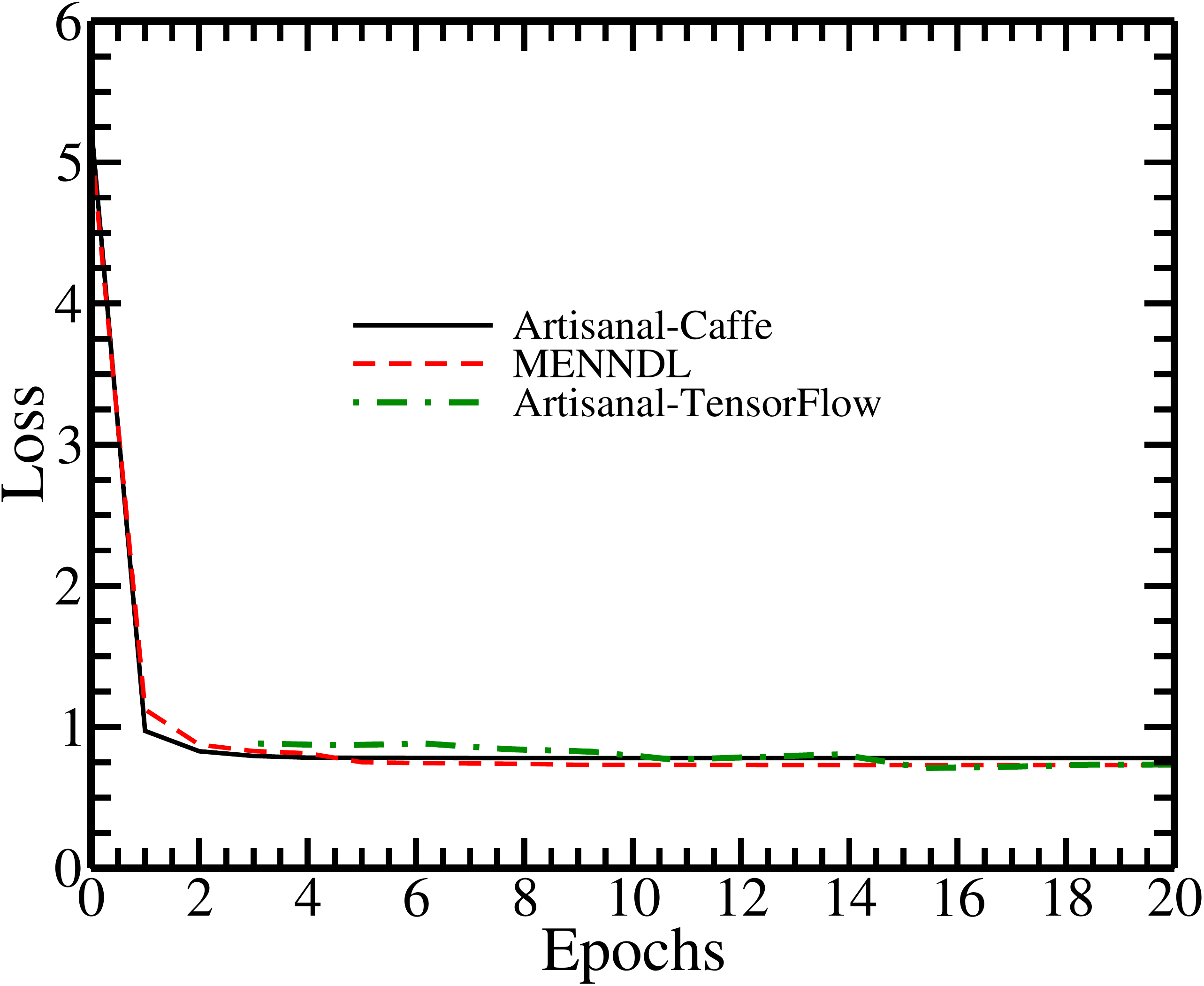}
  \caption{Validation loss}
  \label{fig:loss_comp}
\end{subfigure}
\caption{Validation accuracy and validation loss plots of artisanal-Caffe compared with the best MENNDL (68525) model and artisanal-TensorFlow model.}
\label{fig:acc_loss_comp}
\end{figure}

\begin{figure}[ht!]
\centering
\begin{subfigure}{.5\textwidth}
  \centering
  \includegraphics[width=1.0\linewidth]{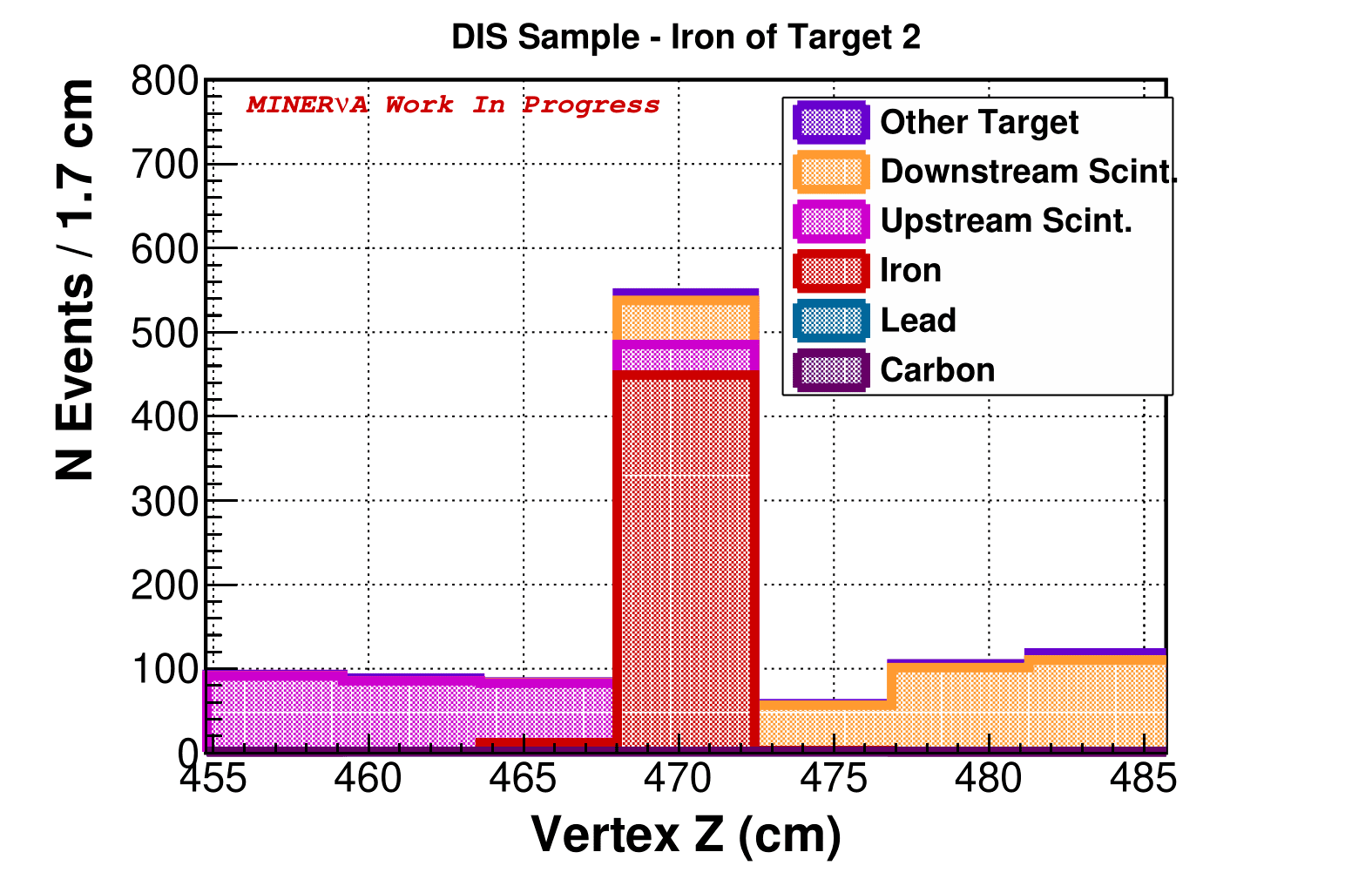}
  \caption{artisanal-TensorFlow}
  \label{fig:stack_TF}
\end{subfigure}%
\begin{subfigure}{.5\textwidth}
  \centering
  \includegraphics[width=1.0\linewidth]{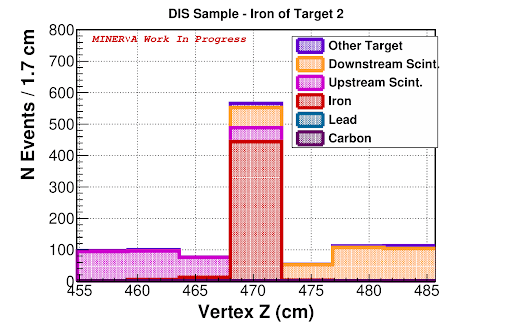}
  \caption{Artisinal-Caffe}
  \label{fig:stack_caffe_artisanal}
\end{subfigure}

\begin{subfigure}{.5\textwidth}
  \centering
  \includegraphics[width=1.0\linewidth]{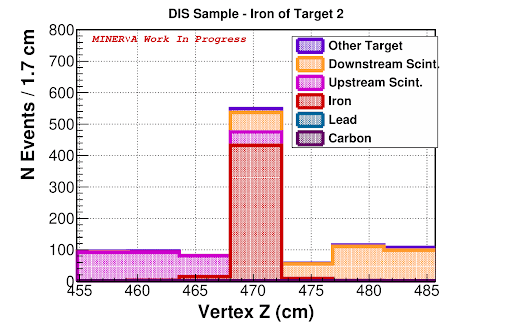}
  \caption{MENNDL}
  \label{fig:stack_caffe_menndl}
\end{subfigure}

\caption{The event distributions for different machine learning models using target 2 of the \minerva detector. Three nuclear targets (iron, lead and carbon) from the \minerva detector were used in this study.  Events reconstructed from the source nuclear target (iron) are represented in red and  other targets, lead and carbon combined, are presented in purple. Distributions of the number of events reconstructed from lead and carbon are shown by blue and indigo colors, respectively. The number of events coming from the plastic scintillator upstream (in this case six planes between targets 1 and 2, closest to target 2) and downstream (in this case six planes between targets 2 and 3, closest to target 2) of the nuclear target are also presented in orange and magenta, respectively. }
\label{fig:DIS_vtzstack}
\end{figure}

The validation accuracy and loss curves of the nine chosen MENNDL models are shown in Fig.~\ref{fig:acc_loss_comb_menndl}. The models are named according to the validation accuracy of the models when they are generated from MENNDL. For example, \textit{MENNDL 57} has validation accuracy of 57\% while generated and evaluated by MENNDL. Among them, we selected the model with highest validation accuracy and compared that to the artisanal-Caffe and artisanal-Tensorflow models. Table~\ref{param_time} shows the number of trainable parameters of each of the MENNDL models and corresponding time of execution. 
\\
Fig.~\ref{fig:acc_loss_comp} shows the comparison of the validation accuracy and validation loss between artisanal-Caffe, arisanal-Tensorflow and MENNDL (MENNDL 68525 as indicated by the yellow line in Fig.~\ref{fig:acc_loss_comb_menndl}) models. The red dashed line corresponds to the MENNDL model whereas the black solid line and green dotted-dashed line correspond to the artisanal-Caffe and artisanal-Tensorflow models, respectively. Figs.~\ref{fig:acc_comp} and~\ref{fig:loss_comp} show the loss and accuracy as the number of epochs increases. The MENNDL model took $\sim$ 20 hours to train whereas the time taken by other two models for the training is more than 24 hours. MENNDL model was automated while the others required human intervention. We can see from the plot that the MENNDL model and artisanal-Caffe model start saturating at the same number of epochs whereas  artisanal-Tensorflow model starts saturating at a higher number of epochs. After ~10 epochs of training, all three models show similar accuracy and loss. Table~\ref{acc_num} lists 
the validation loss and accuracy 
achieved by the three models after reaching the saturation with the total number of adjustable parameters in each model. We also compared the number of parameters of each model, which is shown in the third column of the Table~\ref{acc_num}. 
We see that models found by MENNDL have almost 10 times fewer parameters than the models designed by hand, thus having less expressivity than the more complex artisanal models. This is expected due to the process used by MENDLL, where the fitness of the model is evaluated after a limited number of iterations. However, both artisanal models and MENNDL models demonstrate similar performance, validating the approach used by MENNDL \cite{young2017evolving}.
\\
Next, we test our models where we generate the ML-based predictions of the vertex of each event using those three models followed by the incorporation of those predictions into the \minerva analysis framework. We used a 
simulation sample of about 10\% of the total antineutrino statistics for this purpose.
\\
Fig.~\ref{fig:DIS_vtzstack} shows the event distribution at target 2 with respect to vertex Z using antineutrino simulated data set in Deep Inelastic Scattering (DIS) region. Variable vertex Z is closely related to the actual position of the nuclear targets in the \minerva detector. We use the ML-based prediction for the reconstructed vertex position, where  Figs.~\ref{fig:stack_TF}, ~\ref{fig:stack_caffe_artisanal} and  ~\ref{fig:stack_caffe_artisanal} represent artisanal-TensorFlow, artisanal-Caffe, and MENNDL models, respectively. As target 2 is composed of iron and lead nuclear targets,
the plots are made for the number of events in iron and lead separately. Here we show the event distribution generated in iron. In addition  to the event distributions coming from the nuclear target (iron), number of events coming from the plastic scintillator upstream (in this case six planes between targets 1 and 2, closest to target 2) and downstream (in this case six planes between targets 2 and 3, closest to target 2) of the nuclear target are also presented. All three event distributions are quite similar to each other which signify that all three models provide similar predictions for the interaction vertex.  
We calculate the efficiency and the purity of the sample, where efficiency is the fraction of true events reconstructed. Purity is the fraction of events where the reconstructed value was equal to the underlying true generated value. For the calculation of the efficiency, the numerator is the number of events in the source nuclear target passing the reconstructed sample selection cuts,
the reconstructed and truth DIS cut. The denominator is all true generated CC DIS events in the true fiducial volume with true $\theta_\mu<$ 17$^o$. Purity is related to the efficiency. The purity numerator is same as the efficiency,
divided by the number of events passing the reconstructed sample selection cuts and reconstructed DIS cut.

\begin{figure}[ht]
\centering
  \centering
    \includegraphics[width=0.8\linewidth]{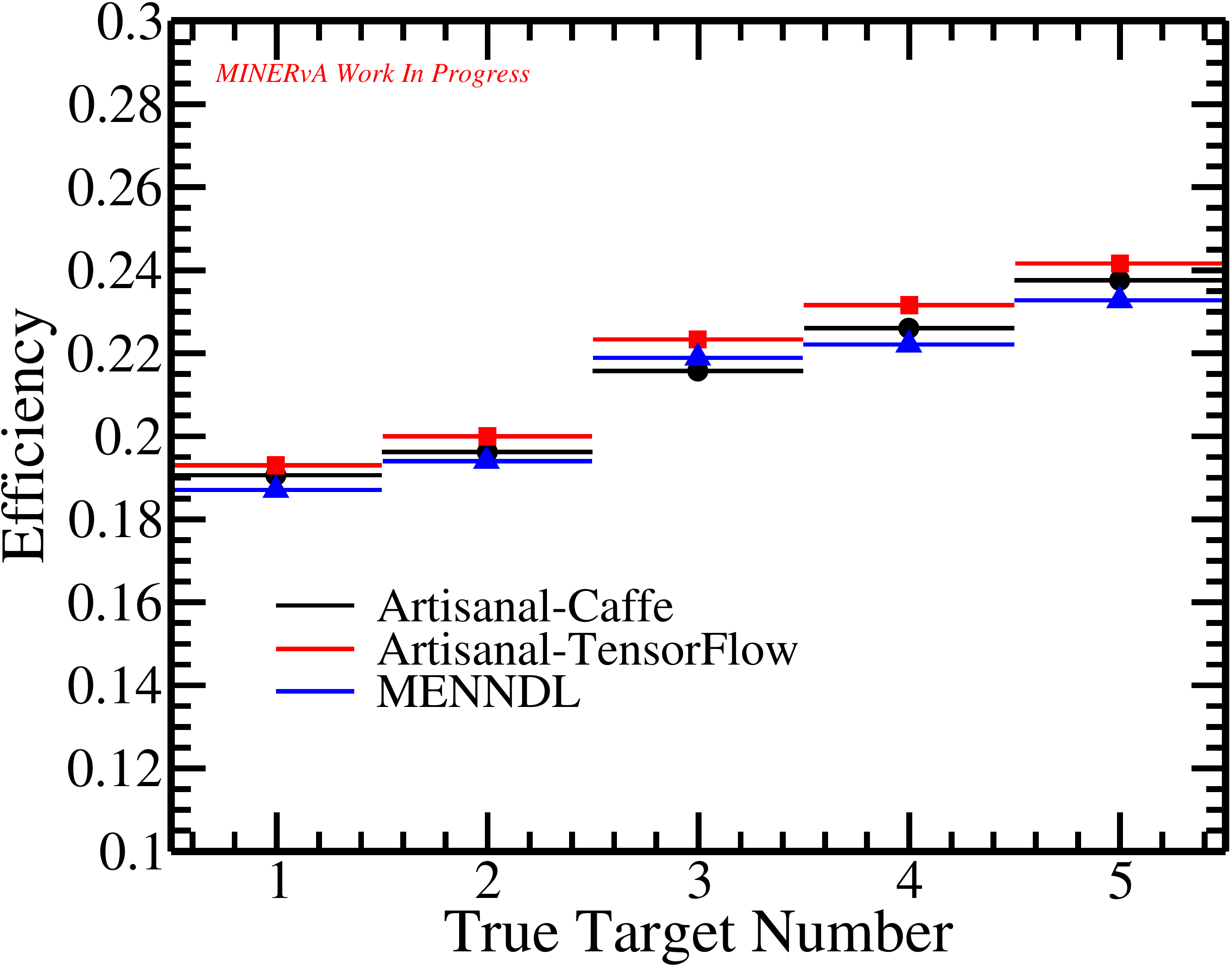}
 \caption{The results for efficiency with respect to five different nuclear targets in MINERvA detector. Different machine
 learning models have been used to make efficiency plots for DIS sample: artisanal-Caffe (black/circle), artisanal-TensorFlow (red/square) and MENNDL (blue/triangle).}.
  \label{fig:DIS_eff}
\end{figure}

\begin{figure}[ht]
\centering
  \centering
  \includegraphics[width=0.8\linewidth]{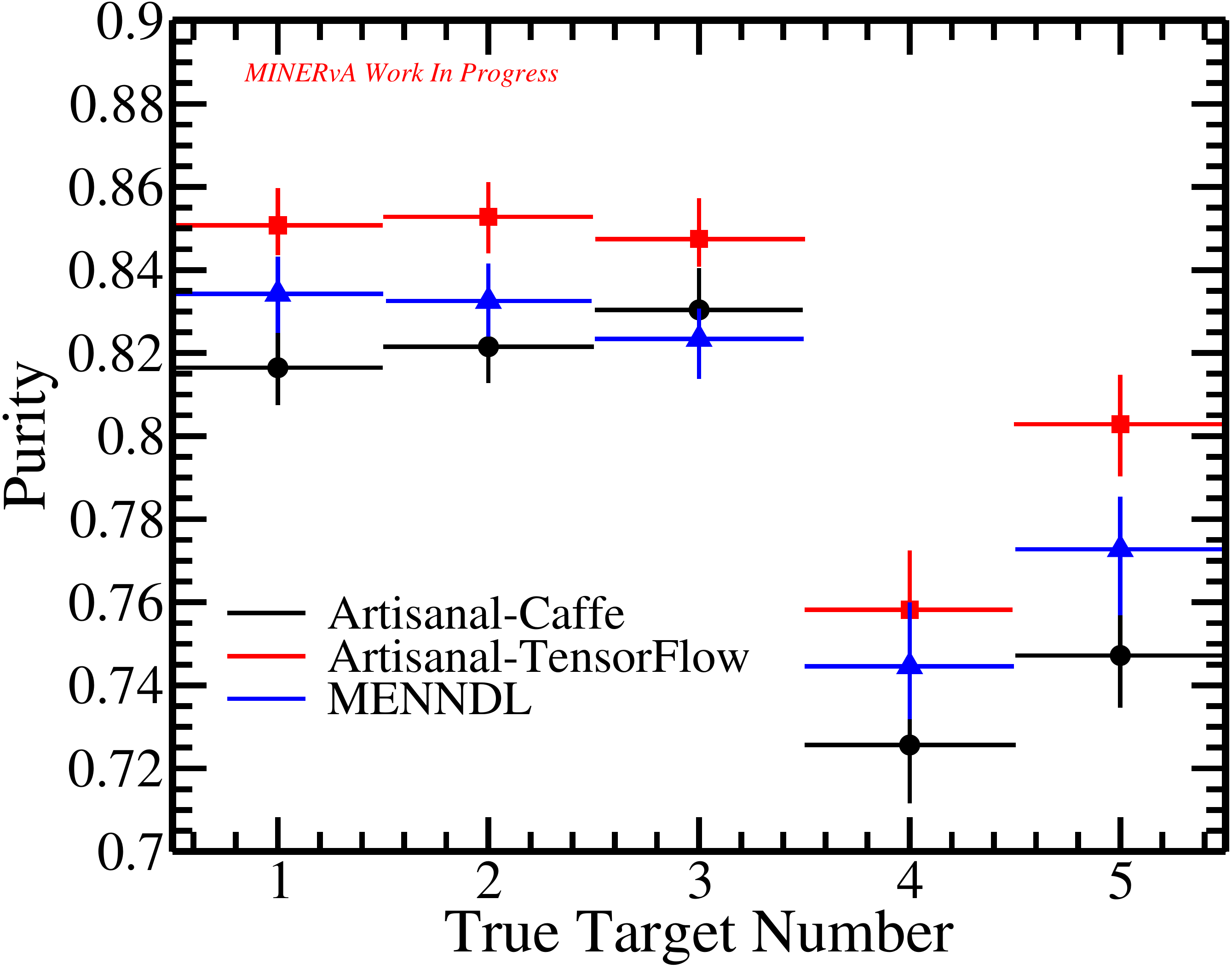}
 
\caption{The results for purity with respect to five different nuclear targets in MINERvA detector. Different machine
 learning models have been used to make Purity plots for DIS sample: artisanal-Caffe (black/circle), artisanal-TensorFlow (red/square) and MENNDL (blue/triangle).}.
  \label{fig:DIS_pur}
\end{figure}

Figs.~\ref{fig:DIS_eff} and ~\ref{fig:DIS_pur} show the efficiency and purity versus true target number.
We compared three models, and the difference between them vary between 2.1\% to 6.8\% and from 1.8\% to 3.2\% for efficiency and purity, respectively (the percent difference is calculated between artisanal-Tensorflow and MENNDL). The plots show that the artisanal-Tensorflow model results in marginally better performance when compared to the other two.



\section{Conclusion}
\label{sec:conclusion}
In this paper, we explored the performance of a ML model obtained from the MENNDL package developed by the ORNL group using MINERvA simulated neutrino data. We further compared the MENNDL model with two successful models which were manually developed by the MINERvA collaboration for improving the vertex reconstruction precision for neutrino interactions. In this work, we trained those models using antineutrino DIS events, generated predictions for the vertex location of the antineutrino DIS events, and compared the performance of the three models in terms of vertex position accuracy at the nuclear target region, efficiency, and purity. In terms of the efficiency with respect to the true target location of DIS events, the differences ranged between 2.1\% to 6.8\% and for the purity the differences varied between 1.8\% to 3.2\% for the three models. We show that the MENNDL generated model performs similarly to the artisanal models. Though both artisanal models have same network structure, artisanal-Tensorflow model shows slightly better performance than artisanal-Caffe model. Primary study indicates that, the internal differences in these two APIs e.g., implication of regularization, SGD e.t.c., are reflected in their performance. This is itself an interesting direction that is beyond the scope of this paper. We also see that models found by MENNDL have almost 10 times fewer parameters than the artisanal models. Hence, MENNDL models have lower capacity than the comparatively more complex artisanal models. However, both models have similar performance. From this we observe that MENNDL favors lower capacity models over more complex ones. These artisanal models were carefully developed, but done manually, and therefore required a significant amount of time of persons with intimate knowledge of the detector's performance. On the other hand, the MENNDL models were generated and evaluated using the package MENNDL where the models are trained over fewer iterations, therefore significantly reducing the amount of specialized human input and time required. 
Furthermore, multiple architectures with similar performance suggests (although not conclusively) that systematics associated with network architecture may be small. 
This technique is quite common in industry for use in automated neural architectures but to the best of our knowledge  this is the first time it has been used to analyze neutrino/antineutrino physics data. In the coming years, the current running and
future neutrino experiments will collect an unprecedented amount of information dense data. As a result, the implementation of a ML algorithm will become necessary to extract the physics embedded in these data, and the MENNDL package will be of immense help in significantly reducing the required human time and  associated time needed to produce an optimized ML model.

\acknowledgments

This document was prepared by members of the \minerva Collaboration using the resources of the Fermi National Accelerator Laboratory (Fermilab), a U.S. Department of Energy, Office of Science, HEP User Facility. Fermilab is managed by Fermi Research Alliance, LLC (FRA), acting under Contract No. DE-AC02-07CH11359.
These resources included support for the \minerva construction project, and support for construction also was granted by the United States National Science Foundation under Award No. PHY-0619727 and by the University of Rochester. Support for participating scientists was provided by NSF and DOE (USA); by CAPES and CNPq (Brazil); by CoNaCyT (Mexico); by Proyecto Basal FB 0821, CONICYT PIA ACT1413, and Fondecyt 3170845 and 11130133 (Chile);  
by CONCYTEC (Consejo Nacional de Ciencia, Tecnolog\'ia e Innovaci\'on Tecnol\'ogica), DGI-PUCP (Direcci\'on de Gesti\'on de la Investigaci\'on  - Pontificia Universidad Cat\'olica del Peru), and VRI-UNI (Vice-Rectorate for Research of National University of Engineering) (Peru); NCN Opus Grant No. 2016/21/B/ST2/01092 (Poland); by Science and Technology Facilities Council (UK); by EU Horizon 2020 Marie Skłodowska-Curie Action; by a Cottrell Postdoctoral Fellowship from the Research Corporation for Scientific Advancement; by an Imperial College London President's PhD Scholarship.  We thank the MINOS Collaboration for use of its near detector data. Finally, we thank the staff of Fermilab for support of the beam line, the detector, and computing infrastructure.

\bibliographystyle{JHEP}
\bibliography{main}
















\end{document}